\documentclass[aps,prd,twocolumn,superscriptaddress,nofootinbib,groupedaddress]{revtex4-2}
 
\pdfoutput=1

\usepackage{amsmath}
\usepackage{amssymb}
\usepackage{calrsfs}
\usepackage{cmupint}
\usepackage{colortbl}
\usepackage{graphicx}
\usepackage{hyperref}
\usepackage{latexsym}
\usepackage{mathtools}
\usepackage{multirow}
\usepackage{rotating}
\usepackage{slashed}
\usepackage{soul}

\newcommand{\ii}{\ensuremath{\mathrm{i}}}
\newcommand{\dd}{\ensuremath{\mathrm{d}}}

\begin{document}


\title{Thermal masses in the Standard Model at three loops}

\author{Mikael Chala}
\email{mikael.chala@ugr.es}
\affiliation{Departamento de F\'isica Te\'orica y del Cosmos, Universidad de Granada, E--18071 Granada, Spain}

\begin{abstract} 
    We compute the Higgs thermal mass and the electroweak and colour Debye mass parameters to full $\mathcal{O}(g^6)$ in dimensional reduction, where $g$ represents a gauge coupling or the top Yukawa. We employ new methods to evaluate previously unknown mixed-signature sum-integrals.
\end{abstract}

\maketitle


%
\section{Introduction}
Elementary particles acquire thermal effective masses $m_\text{eff}$ in their interaction with a thermal bath. They describe changes on the corresponding field's response to a disturbance within the plasma. The case of the photon mass is very illustrative: Inserting a charge into a plasma makes the mobile charges rearrange, partially neutralizing its field, $\phi(r)\sim e^{-m_\text{eff}r}/r$. This phenomenon applies to gauge theories in general and it is known as \textit{Debye screening}.

From the quantum field theory (QFT) perspective, thermal masses encode quantum corrections to two-point particle propagators. Three relevant energy ($E$) scales characterize these corrections: the hard ($E\sim \pi T$), soft ($E\sim gT$) and ultrasoft ($E\sim g^2 T$) scales. The first can be handled perturbatively, while the latter is genuinely non-perturbative~\cite{Linde:1980ts}. The dimensional-reduction program~\cite{Ginsparg:1980ef,Appelquist:1981vg,Braaten:1995cm,Kajantie:1995dw} exploits this separation of scales to integrate out the hard scale, encoding its effects on the Wilson coefficients of an effective field theory (EFT). This EFT is 3-dimensional (3D), reflecting that QFT at finite temperature ($T$) is equivalent to regular (Euclidean) QFT where the time dimension is compactified in a circle of circumference $1/T$. Fourier expansion around the com-
pact time direction gives Matsubara modes~\cite{Matsubara:1955ws}. Hard matching integrates out the nonzero modes and includes the hard-momentum contributions of the bosonic zero modes. It turns out that fermionic fields, which obey anti-periodic boundary conditions on the time direction, have no zero modes; so the EFT involves only bosons. It is therefore an excellent setup to handle soft and, most importantly, ultrasoft quantum corrections on the lattice~\cite{Farakos:1994xh,Kajantie:1995kf,Andersen:2017ika,Niemi:2018asa,Gould:2019qek,Kainulainen:2019kyp,Niemi:2020hto,Gould:2021dzl,Gould:2022ran,Niemi:2024axp,Gould:2024chm,Niemi:2024vzw,Annala:2025aci,Chala:2025gif}.

Accurately determining the hard contributions to thermal masses has been subject of intense research for decades. Let us assume the power counting $\mu^2,\lambda,y_t^2,g_1^2,g_2^2,g_s^2\sim g^2$, representing the Higgs quadratic and quartic terms, the top Yukawa and the SM $U(1)$, $SU(2)$ and $SU(3)$ gauge couplings. Thus, the $\mathcal{O}(g^4)$ Higgs and Standard Model (SM) Debye masses were computed in \cite{Gynther:2005dj,Gynther:2005av}; see also \cite{Ekstedt:2022bff}. (Previous works~\cite{Farakos:1994kx,Kajantie:1995dw} assumed $g_1^2\sim g^3$). $\mathcal{O}(g^6)$ was achieved for the QCD Debye mass in pure Yang-Mills in \cite{Ghisoiu:2015uza}; the full QCD computation, including quarks (but dismissing electroweak couplings), was addressed in \cite{Moller:2012chx}, though certain sum-integrals remain unevaluated; see also \cite{Laine:2019uua} for previous results at $\mathcal{O}(g^4)$.

The main motivation behind these works is that thermal masses are instrumental to accurately predict a variety of observables, including the pressure~\cite{Braaten:1995jr,Kajantie:2002wa}, spatial string tension~\cite{Laine:2005ai,Moller:2012chx}, static heavy-quark screening~\cite{Burnier:2009bk}, jet broadening~\cite{Caron-Huot:2008zna,Panero:2013pla} and particle annihilation rates~\cite{Kim:2016kxt}, among others. Moreover, recent studies demonstrate the importance of $\mathcal{O}(g^6)$ hard matching for precision studies of thermal phase transitions within dimensionally reduced EFTs~\cite{Chala:2024xll,Bernardo:2025vkz,Chala:2025aiz,Chala:2025oul,Biekotter:2025npc,Chala:2025cya,Fuentes-Martin:2026bhr,Bernardo:2026nyq,Bernardo:2026whs,Chakrabortty:2026swu,Bandyopadhyay:2026nrv}.

In this \textit{Letter}, we compute the Higgs and the electroweak and colour Debye masses through $\mathcal{O}(g^6)$ in the full SM.

\section{Conventions}
We use the Minkowski SM Lagrangian in the background-field gauge, where gauge bosons are split into background ($X$) and quantum ($Q_X$) fluctuations, with $X=B,W,G$. The Lagrangian reads
\begin{align}
 \mathcal{L}_\text{SM} &= \mathcal{L}_\text{fermion} + \mathcal{L}_\text{Higgs} +(\mathcal{L}_\text{Yukawa}+\text{h.c.})\\
 &+ \mathcal{L}_\text{gauge} + \mathcal{L}_\text{ghost} + \mathcal{L}_\text{gauge-fixing}\,.\nonumber
\end{align}
with
\begin{align}
    \mathcal{L}_\text{Higgs} &= (D_\mu\phi)^\dagger (D^\mu\phi) -\mu^2|\phi|^2 -\frac{\lambda}{2}|\phi|^4\,, \nonumber\\
    \mathcal{L}_\text{fermion} &= \overline{q}\ii\slashed{D}q+\overline{l}\ii\slashed{D} l+\overline{u}\ii\slashed{D}u+\overline{d}\ii\slashed{D}d+\overline{e}\ii\slashed{D}e\,,\nonumber\\
    \mathcal{L}_\text{Yukawa} &= - (\overline{q} \widetilde{\phi} Y_u u + \overline{q}\phi Y_d d+\overline{l}\phi Y_e e)\,,\\
    \mathcal{L}_\text{gauge} &= -\frac{1}{4} G_{\mu\nu}^A G^{A\mu\nu} - \frac{1}{4}W^I_{\mu\nu}W^{I\,\mu\nu}-\frac{1}{4}B_{\mu\nu}B^{\mu\nu}\,, \\
    \mathcal{L}_\text{ghost} &= -\overline{c_G}^A \bigl(\overline{D}^\mu D_{\mu}c_G\bigr)^A -\overline{c_W}^I \bigl(\overline{D}^\mu D_{\mu}c_W\bigr)^I\,,\\
    \mathcal{L}_\text{gauge-fixing} &= -\frac{1}{2\xi_G} \bigl(\overline{D}^\mu Q_{G\mu}\bigr)^A\bigl(\overline{D}^\nu Q_{G\nu}\bigr)^A\nonumber\\
    &-\frac{1}{2\xi_W}\bigl(\overline{D}^\mu Q_{W\mu}\bigr)^I \bigl(\overline{D}^\nu Q_{W\nu}\bigr)^I -\frac{1}{2\xi_B} \bigl(\partial^\mu Q_{B\mu}\bigr)^2\,.
\end{align}

Here $\overline{D}$ contains only background fields, while $D$ and the field strengths contain background plus quantum fields. The Abelian ghost is omitted because it is free. We
use the minus-sign convention for the covariant derivative.

Our only simplifying assumption is $Y_e=Y_d=0$ and $Y_u=\text{diag}(y_t,0,0)$.

\begin{figure*}[t]
    \includegraphics[width=2\columnwidth]{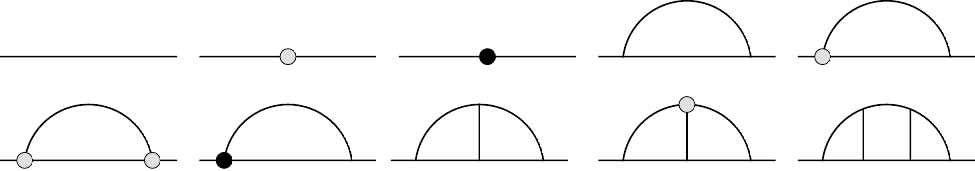}
    \caption{\it Schematic representation of Feynman diagrams contributing to two-point correlators. Gray and black dots stand for one- and two-loop counterterms, respectively. These are taken from Refs.~\cite{Luo:2002ey,Bednyakov:2012rb}. We deliberately omit the tree-level diagram with a three-loop counterterm because it is $\mathcal{O}(g^8)$. Gauge-fixing counterterms only affect off-shell calculations.}\label{fig:diagrams}
\end{figure*}

The relevant off-shell Euclidean 3D EFT Lagrangian reads
\begin{widetext}
\begin{align}\label{eq:3deft}
    \mathcal L_3 &= m_\phi^2 \phi^\dagger \phi + \frac{1}{2}m_{B_0}^2 B_0^2 + \frac{1}{2}m_{W_0}^2 W_0^I W_0^I + \frac{1}{2}m_{G_0}^2 G_0^A G_0^A\nonumber\\
    & + k_\phi D_i\phi^\dagger D_i\phi + \frac{1}{2}k_{B_0} D_iB_0 D_i B_0 + \frac{1}{2}k_{W_0} D_iW_0^I D_i W_0^I+ \frac{1}{2}k_{G_0} D_i G_0^A D_i G_0^A\nonumber\\
    &+ r_{\phi^2 D^4}D_i^2\phi^\dagger D_i^2\phi + \frac{1}{2} r_{B_0^2 D^4} D_i^2 B_0 D_i^2 B_0 + \frac{1}{2} r_{W_0^2 D^4} D_i^2 W_0^I D_i^2 W_0^I + \frac{1}{2} r_{G_0^2 D^4} D_i^2 G_0^A D_i^2 G_0^A\,.
\end{align}
\end{widetext}
Upon canonical normalization and removing the four-derivative operators using field redefinitions, we have $m_\phi^2\to m_\phi^2/k_\phi+r_{\phi^2 D^4}m_\phi^4$, and likewise for the Debye masses, conveniently truncated to $\mathcal{O}(g^6)$. This requires computing (off-shell) masses up to three loops, kinetic terms up to two loops and four-derivative operators at one loop. Schematic diagrams are shown in Fig.~\ref{fig:diagrams}.

These masses run first at two loops. The corresponding counterterms can be found in \cite{Kajantie:1995dw,Kajantie:1997tt,Ekstedt:2022bff,Chala:2025cya}.

\section{Methodology}
We match the 3D coefficients in Eq.~\eqref{eq:3deft} diagrammatically, equating the hard-region expansion of static two-point correlators (where external momenta $\vec{p}$ and the Higgs mass parameter $\sqrt{\mu^2}$ are much smaller than loop momenta $P,Q,R$) to tree-level counterpart in the EFT. We compute the relevant amplitudes using dedicated routines based on \texttt{FeynRules}~\cite{Alloul:2013bka}, \texttt{FeynArts}~\cite{Hahn:2000kx} and \texttt{FeynCalc}~\cite{Shtabovenko:2023idz}. We use dimensional regularization with dimension $d=3-2\epsilon$ and fix $\xi_B=\xi_W=\xi_G=1$.

Since infinitely many Matsubara modes run in loops, Wilson coefficients depend on sum-integrals. At one loop we have
%
%
\begin{equation}
      \mathcal{I}_{\alpha,s}^{\beta}
      \equiv \sumint_{P,s}\frac{P_0^\beta}{(P^2)^\alpha}
      \equiv
      \widetilde{\mu}^{2\epsilon}
      T\sum_{n\in\mathbb Z}
      \int\frac{\mathrm d^d p}{(2\pi)^d}
      \frac{\omega_{n}^{\beta}}
           {(\vec p^{\,2}+\omega_{n}^{2})^\alpha}\,,
\end{equation}
where $\omega_n=2\pi T(n+s/2)$ and $s=0 (1)$ for bosons (fermions), and $\widetilde{\mu}^2=e^{\gamma_E}\overline{\mu}^2/(4\pi)$ with $\overline{\mu}$ the $\overline{\text{MS}}$ renormalization scale. Odd frequency numerators vanish; even ones reduce to numerator-free one-loop integrals $\mathcal{I}_{\alpha,s}\equiv \mathcal{I}_{\alpha,s}^0$ with Gamma-function prefactors. The latter are
\begin{align}
    \mathcal{I}_{\alpha,0} &= \frac{2T}{(4\pi)^{d/2}} \frac{\Gamma\left(\alpha-\frac{d}{2}\right)}{\Gamma(\alpha)} (2\pi T)^{d-2\alpha}\, \widetilde{\mu}^{2\epsilon}\, \zeta(2\alpha-d)\,, \nonumber\\
    \mathcal{I}_{\alpha,1} &= \left(2^{2\alpha-d}-1\right)\mathcal{I}_{\alpha,0}\,.
\end{align}
Two-loop sum-integrals depend on at most three propagators ($P^2$, $Q^2$ and $(P-Q)^2$) and two temporal modes ($P_0$ and $Q_0$). Following the notation above, we denote a generic such sum-integral as $\mathcal{I}_{\alpha_1\alpha_2\alpha_3,s_1s_2}^{e_1,e_2}$. Any such sum-integral can be factorized into products of one-loop sum-integrals using integration-by-parts (IBP) methods; see \cite{Nishimura:2012ee,Davydychev:2023jto,Gil:2026cqz}. For example:
\begin{align}
    \mathcal{I}_{211,01}^{0,0} &= \frac{1}{(d-5)(d-2)}[(\mathcal{I}_{2,1})^2 -2\mathcal{I}_{2,1}\mathcal{I}_{2,0}]\,.
\end{align}

Likewise, we adopt the following conventions for three-loop sum-integrals:
\begin{equation}
   \mathcal{I}_{\alpha_1,\cdots\alpha_6,s_1s_2s_3}^{e_1e_2e_3} =\sumint_{PQR} \frac{P_0^{e_1} Q_0^{e_2} R_0^{e_3}}{P^{2\alpha_1} \cdots (Q-R)^{2\alpha_6}}\,.
\end{equation}
These, in general, do not factorize, but can often be expressed in terms of simpler sum-integrals using IBP relations. To this aim, we use \texttt{SIRENA}~\cite{Gil:2026cqz}. This code allows one also to specify certain integrals under the \texttt{priority} flag, that are kept as master integrals even before simpler sum-integrals. For this calculation, we use $\mathcal{I}_{111110,000}^{000}$~\cite{Andersen:2008bz,Schroder:2012hm}, $\mathcal{I}_{211110,000}^{020}$~\cite{Ghisoiu:2012kn}, $\mathcal{I}_{31111-2,000}^{000}$~\cite{Ghisoiu:2012yk} and $\mathcal{I}_{210011,000}^{000}$~\cite{Moller:2010xw,Ghisoiu:2012kn}, together with the mixed-statistics sum-integrals in Tab.~\ref{tab:master-coefficients}, that we obtain here for the first time. The values $a,b,c$ are the numerators in
\begin{align}
    \mathcal I=\frac{1}{(4\pi)^4}\left[\frac{a}{\epsilon^2}+\frac{b}{\epsilon}+c+O(\epsilon)\right]\,.
\end{align}
obtained at $T=\overline{\mu}=1$. The corresponding results at arbitrary $\overline{\mu}$ can be trivially obtained as
\begin{align}
    \mathcal I = \frac{T^2}{(4\pi)^4}\left[\frac{a}{\epsilon^2}+\frac{b+6a\ell}{\epsilon}+c+6b\ell+18a\ell^2+\mathcal{O}(\epsilon)\right]
\end{align}
where $\ell=\log{\dfrac{\overline{\mu}}{T}}$.
\begin{table}[t]
  \centering
  \resizebox{\columnwidth}{!}{
      \begin{tabular}{lrrr}
      \hline\\[-0.2cm]
       & $a$ & $b$ & $c$ \\[0.1cm]
      \hline\\[-0.2cm]
      $\mathcal{I}_{111110,001}^{000}$
      & $-0.12500(2)$ & $0.154(2)$ & $-0.97(2)$ \\[0.1cm]
      $\mathcal{I}_{111110,100}^{000}$
      & $0$ & $-0.0738(1)$ & $-0.113(2)$ \\[0.1cm]
      $\mathcal{I}_{210011,001}^{000}$
      & $0$ & $0.0369286(3)$ & $-0.001838(8)$ \\[0.1cm]
      $\mathcal{I}_{210011,100}^{000}$
      & $0.062495(6)$ & $0.0172(3)$ & $0.314(4)$ \\[0.1cm]
      $\mathcal{I}_{21111-1,100}^{000}$
      & $0.0208335(2)$ & $0.058266(7)$ & $0.0427(3)$ \\[0.1cm]
      $\mathcal{I}_{211110,001}^{002}$
      & $-0.005207(2)$ & $0.0081(1)$ & $0.146(2)$ \\[0.1cm]
      $\mathcal{I}_{211110,101}^{011}$
      & $0.00781251(1)$ & $-0.018159(3)$ & $-0.00241(5)$ \\[0.1cm]
      $\mathcal{I}_{211110,001}^{020}$
      & $-0.010433(9)$ & $0.1094(2)$ & $-0.436(3)$ \\[0.1cm]
      $\mathcal{I}_{211110,011}^{020}$
      & $-0.02604158(4)$ & $0.035819(5)$ & $0.00783(8)$ \\[0.1cm]
      $\mathcal{I}_{211110,101}^{101}$
      & $0.0234375(2)$ & $0.00023(2)$ & $0.0212(3)$ \\[0.1cm]
      $\mathcal{I}_{31111-2,001}^{000}$
      & $-0.05556(1)$ & $0.1182(5)$ & $-0.509(4)$ \\[0.1cm]
      $\mathcal{I}_{31111-2,011}^{000}$
      & $0.02777(2)$ & $-0.2646(4)$ & $0.790(5)$ \\[0.1cm]
      \hline
      \end{tabular}
  }
  \caption{$\epsilon$-expansion of different sum-integrals. Parentheses give statistical Monte Carlo errors.}\label{tab:master-coefficients}
  \end{table}

Two other sum-integrals appear in our calculation, but they can be trivially traded for previous ones:
\begin{align}
    \mathcal I_{111110,011}^{000}&=\left(2^{10-3d}-1\right)\mathcal I_{111110,000}^{000}\nonumber\\
    &-2\mathcal I_{111110,001}^{000}-4\mathcal I_{111110,100}^{000}\,,\\
    \mathcal I_{210011,110}^{000} &=
    \left(2^{10-3d}-1\right)\mathcal I_{210011,000}^{000}\nonumber\\
    &-3\mathcal I_{210011,001}^{000}-3\mathcal I_{210011,100}^{000}.
\end{align}
The rational coefficients of all the aforementioned non-factorized masters are regular at $d =3$ in the matching coefficients. No positive-order $\epsilon$ coefficients of the masters are therefore needed.

\section{New sum-integral evaluations}
We have numerically evaluated the sum-integrals in Tab.~\ref{tab:master-coefficients} following a combination of Schwinger parametrization of propagators~\cite{Schwinger:1951nm}, sector decomposition~\cite{Binoth:2000ps,Harnett:2024nny} and partial Poisson summation~\cite{Niedermayer:2016ilf,Nishimura:2018wla}.

The aim is to express any given sum-integral as a combination of expressions of the form
\begin{equation}
    I(\epsilon) = \int_0^1 \dd t\, t^{\epsilon-1} F(t)\,,
\end{equation}
with $F$ continuous differentiable in $0\leq t\leq1$. This way, the $1/\epsilon$ poles can be subtracted following
\begin{align}\label{eq:endpoint}
    I(\epsilon) &= F(0)\int_0^1 \dd t\, t^{\epsilon-1} + \int_0^1 \dd t\, t^{\epsilon-1}[F(t)-F(0)]\nonumber\\
    &=\frac{F(0)}{\epsilon} + \text{finite}\,.
\end{align}
Given that $F(t)-F(0)$ is $O(t)$ as $t\to0$, the subtracted integrand behaves as $t^\epsilon$ and the remainder is therefore finite near $\epsilon=0$ and can be evaluated numerically. 

Schwinger parametrization amounts to replace Feynman propagators by integrals over real parameters $s_i$ following
\begin{equation}
    \frac{1}{P^{2\alpha}} = \frac{1}{\Gamma(\alpha)}\int_0^\infty \dd s_i\, s_i ^{\alpha-1} e^{-s_i P^2}\,.
\end{equation}

After introducing dimensionless Schwinger parameters $x_i=(2\pi T)^2s_i$, sector decomposition splits each integration range into small ($x_i<1$) and large ($x_i>1$) regions. The corresponding propagators are called \textit{small lines} and \textit{large lines}, respectively. Partial Poisson summation uses the identity
\begin{equation}
    \sum_{n\in\mathbb{Z}} e^{-t(n+\delta)^2} = \sqrt{\frac{\pi}{t}}\sum_{w\in\mathbb{Z}} e^{2\pi iw\delta}e^{-\pi^2w^2/t}
\end{equation}
to swap sums over Matsubara frequencies for sums over winding modes; only in small lines. 

All this becomes clearer with a one-loop example.

Let us consider the fermionic sum-integral
\begin{align}
    J(\epsilon) \equiv \mathcal{I}_{2,1} = \widetilde{\mu}^{2\epsilon} T \sum_{n\in\mathbb{Z}}\int \frac{\dd^d p}{(2\pi)^d}\frac{1}{(\vec{p}^2+\omega_n^2)^2}\,,
\end{align}
so $\omega_n = 2\pi T(n+1/2)$.

Introducing Schwinger parameters, we obtain
\begin{equation}
    J(\epsilon) = \widetilde{\mu}^{2\epsilon} T \sum_{n\in\mathbb{Z}}\int_0^\infty \dd s\, s\, e^{-s \omega_n^2}\int \frac{\dd^d p}{(2\pi)^d}\, e^{-s\vec{p}^2}\,.
\end{equation}
The integral over $d$-dimensional momenta can be trivially solved. Therefore, performing the change of variables $x=(2\pi T)^2 s$, we obtain
\begin{equation}\label{eq:reducedintegral}
    J(\epsilon) = \frac{1}{16\pi^2} \left(\frac{\widetilde{\mu}^2}{\pi T^2}\right)^\epsilon \overbrace{\frac{1}{\sqrt{\pi}} \int_0^\infty \dd x\, x^{\epsilon-\frac{1}{2}}\sum_{n\in\mathbb{Z}} e^{-x(n+\frac{1}{2})^2}}^{I(\epsilon)}\,.
\end{equation}
%

The factor multiplying $I(\epsilon)$ is completely regular at $\epsilon=0$. So we focus on $I(\epsilon)$.

We do the change of variables $x=t$ for $0<x<1$ and $x=1/t$ for $1<x<\infty$. This way, the integral takes the form
\begin{align}
    I(\epsilon) &= \frac{1}{\sqrt{\pi}} \int_0^1 \dd t\, t^{\epsilon-\frac{1}{2}} \sum_{n\in\mathbb{Z}} e^{-t(n+\frac{1}{2})^2}\nonumber\\
    &+ \frac{1}{\sqrt{\pi}} \int_0^1 \dd t\, t^{-\epsilon-\frac{3}{2}} \sum_{n\in\mathbb{Z}} e^{-(n+\frac{1}{2})^2/t}\,.
\end{align}
The second sum is exponentially suppressed at $t\to0$. 
The first, however, is not, because the sum behaves badly at $t\to 0$. So we do Poisson summation in this term only, obtaining
\begin{align}
    I(\epsilon) = \int_0^1 \dd t\, t^{\epsilon-1} F_0(t) + \int_0^1 \dd t\, t^{-\epsilon-\frac{3}{2}} F_1(t)\,,
\end{align}
with $F_0(t) = 1 + 2\sum_{w=1}^{\infty} (-1)^w e^{-\pi^2 w^2/t}$ and where $F_1(t) = \frac{2}{\sqrt{\pi}} \sum_{n=0}^\infty e^{-(n+\frac{1}{2})^2/t}$.
Since $F_0(t)-1$ and $F_1(t)$ vanish exponentially as $t\to0$, the pole comes entirely from $F_0(0)=1$. Applying Eq.~\eqref{eq:endpoint} gives
\begin{align}
    I(\epsilon) = \frac{1}{\epsilon} + \int_0^1\dd t \, t^{\epsilon-1} [F_0(t)-1] + \int_0^1\dd t \, t^{-\epsilon-\frac{3}{2}} F_1(t)\,. 
\end{align}
Both remaining integrals are finite near $\epsilon=0$. Their expansion yields convergent integrals for the Laurent coefficients, which can be evaluated numerically upon truncating the sums to a desired finite order, depending on the precision to achieve.

This method can be applied equally well at higher loops. The main difference is the proliferation of sectors. For example, a two-loop sum-integral with three distinct propagator denominators has $3!=6$ orderings of its Schwinger parameters, each of which can in turn be split into four sectors: $1>x_0>x_1>x_2$, $x_0>1>x_1>x_2$, $x_0>x_1>1>x_2$ and $x_0>x_1>x_2>1$. The corresponding changes of variables moving the limits of integration from $[0,\infty]$ to $[0,1]$ are then
\begin{align}
    (x_0,x_1,x_2)&=(t_0,t_0t_1,t_0t_1t_2)\,,\\
    (x_0,x_1,x_2)&=(\frac{1}{t_0},t_1,t_1 t_2)\,,\\
    (x_0,x_1,x_2)&=(\frac{1}{t_0t_1},\frac{1}{t_1},t_2)\,,\\
    (x_0,x_1,x_2)&=(\frac{1}{t_0t_1t_2},\frac{1}{t_1t_2},\frac{1}{t_2})\,;
\end{align}
respectively. Moreover, we change basis before partial Poisson summation when the large-line momenta cannot all be expressed using just $r$ of the loop variables, where $r$ is the number of linearly independent momentum combinations among the large lines. We choose a lattice-preserving basis where they can; retaining those $r$ frequency sums and Poisson-transforming only the remaining $L-r$, where $L$ is the number of loops.
Endpoint subtraction is then performed successively in the singular variables, including higher Taylor terms when required. After subtraction, the regular factors are expanded in \(\epsilon\), leaving finite integrals for the Laurent coefficients.

Despite building on earlier work~\cite{Niedermayer:2016ilf,Nishimura:2018wla,Cavalcanti2021,Harnett:2024nny},
this approach provides, to the best of our knowledge,
the first numerical determination of Laurent coefficients
of mixed bosonic--fermionic three-loop thermal sum-integrals
by combining ordered Schwinger-parameter sectors,
sector-dependent partial Poisson summation, and analytic
endpoint subtraction. Although not needed for the present calculation, the same framework provides a systematic route to higher orders in the $\epsilon$ expansion.

\section{Results}
The ancillary file \texttt{matching.txt} gives the \textit{bare} hard two-point coefficients, including four-dimensional parameter counterterms, in terms of one-loop integrals and  three-loop masters. It precedes the final canonical normalization and 3D counterterm subtraction. 

For readability only, we provide here the numerical expressions for the \textit{renormalized} masses $m_{\phi}^{2}$ and $m_{G_0}^2$ through $\mathcal{O}(g^6)$ in the limit $g_1,\lambda,\mu^2\to 0$, adopting the power counting $g_2^2\sim g^3$, and fixing the 3D renormalization scale to $\mu_3=\pi T$:
\begin{widetext}
\begin{align}
    \frac{m_{\phi}^2}{T^2}&\simeq 0.19 g_2^2+0.25 y_t^2+\bigl(-0.015+0.025\ell\bigr)g_s^2y_t^2+\bigl(-0.0093+0.0047\ell\bigr)y_t^4 +\bigl(-0.017+0.0071\ell\bigr)g_2^2y_t^2\nonumber\\
    &+\bigl(-0.0012+0.0040\ell\bigr)g_2^4 +\bigl(0.0012-0.00075\ell+0.0024\ell^2\bigr)g_s^4y_t^2+\bigl(0.00016-0.0011\ell-0.00024\ell^2\bigr)g_s^2y_t^4 \nonumber\\
    &+\bigl(-0.00086+0.0030\ell-0.00099\ell^2\bigr)y_t^6\,,
\end{align}
\begin{align}
      \frac{m_{G_0}^2}{T^2}
      &\simeq 2g_s^2
       +\bigl(-0.42+0.18\ell\bigr)g_s^4
       -0.0063g_s^2y_t^2-0.014g_s^2g_2^2 \nonumber\\
      &+\bigl(0.087-0.071\ell+0.016\ell^2\bigr)g_s^6
       +\bigl(0.0020-0.00088\ell\bigr)g_s^4y_t^2
       +\bigl(-0.00037+0.00036\ell\bigr)g_s^2y_t^4\,.
  \end{align}
\end{widetext}
where again $\ell=\log{\frac{\overline{\mu}}{T}}$. 

The results in \texttt{matching.txt} reduce to previous results in the literature in the corresponding limits. (We could not cross-check the partial $\mathcal{O}(g^6)$ results in \cite{Moller:2012chx} because we use very different bases of sum-integrals.) Moreover, they pass several non-trivial consistency checks; most importantly, the renormalized masses (i.e. $m_{\phi}^2-\delta m_{\phi}^2$, where $\delta m_\phi^2$ represents the 3D counterterm, and likewise for the Debye masses) are completely pole free.

\begin{figure}[t]
    \includegraphics[width=\columnwidth]{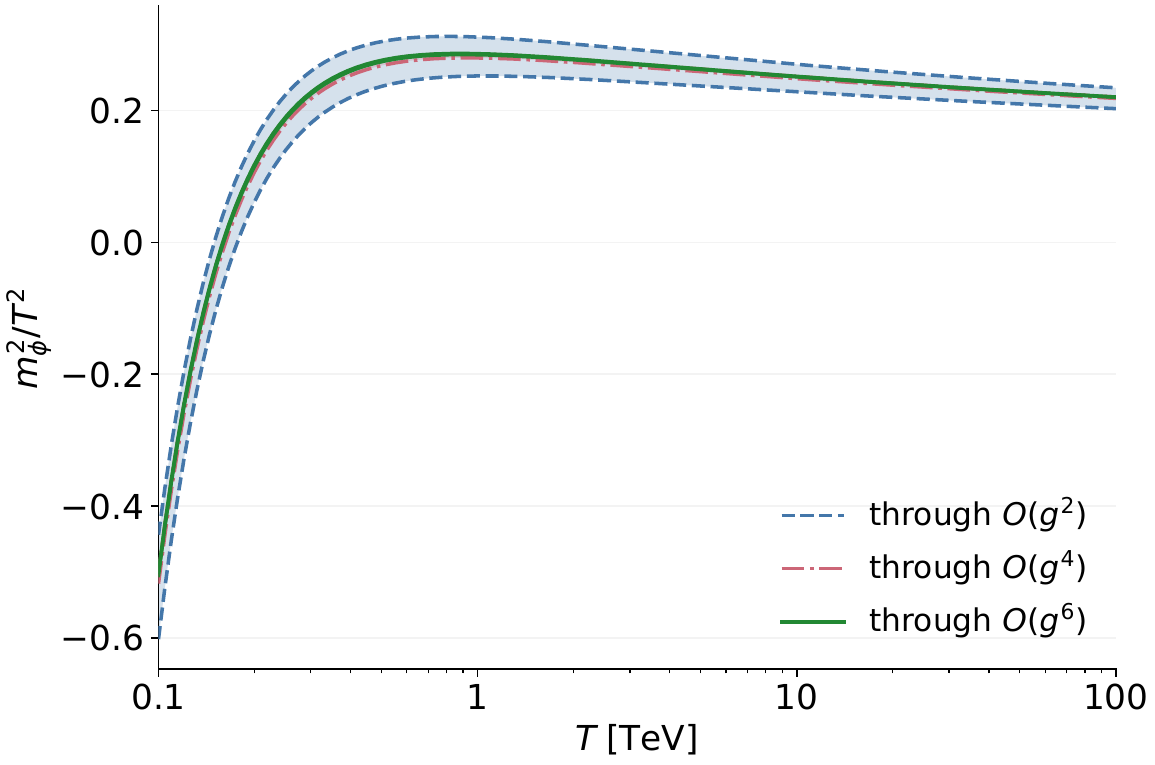}
    \caption{\it Scale dependence of the renormalized Higgs mass at different orders, for $\overline{\mu}\in [\pi T/4, 4\pi T]$. The 3D scale is fixed to $\mu_3=\pi T$. SM inputs are taken from interpolation formulas (55)-(60) in \cite{Buttazzo:2013uya}. Two-loop running of 4D parameters is performed numerically in all cases.}\label{fig:scaledependence1}
\end{figure}

\begin{figure}[b]
    \includegraphics[width=\columnwidth]{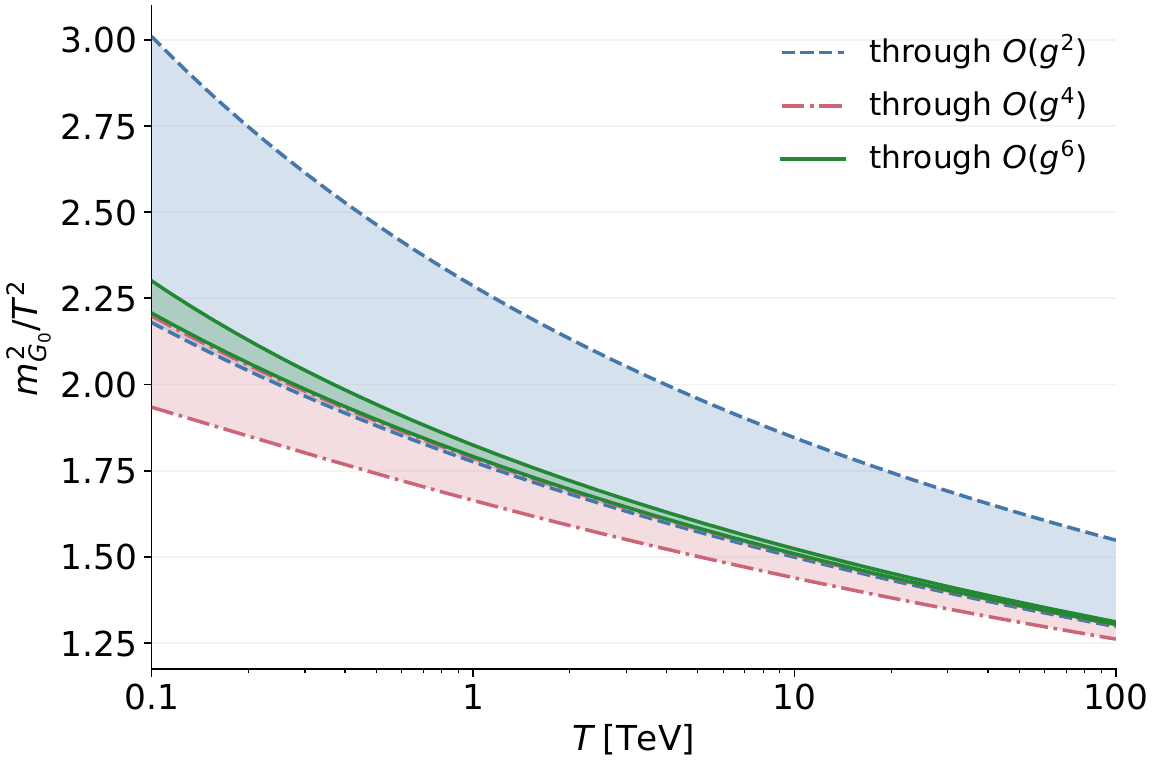}
    \caption{\it Same as Fig.~\eqref{fig:scaledependence1} but for the colour Debye mass.}\label{fig:scaledependence2}
\end{figure}

For example, the $y_t^4 g_s^2$ piece in the (bare) colour Debye mass is
\begin{widetext}
\begin{align}\label{eq:cancellations}
      m_{G_0}^2\bigg|_{y_t^4 g_s^2}
      %
      &= y_t^4 g_s^2\bigg\{
      (d-3)\bigg[
      -12\mathcal{I}_{210011,100}^{000}
      +24\mathcal{I}_{210011,110}^{000}
      -24(d-4)\left(
      \mathcal{I}_{211110,101}^{011}
      +\mathcal{I}_{211110,101}^{101}\right)
      \bigg]
      -\frac{3T^2}{256\pi^4\epsilon}
      \bigg\}+\text{finite}\,,
      %
\end{align}
\end{widetext}
where the explicit pole comes from factorizable sum-integrals.

Following Tab.~\ref{tab:master-coefficients}, double poles cancel trivially. Numerically, the single pole ensuing from the non-factorized three-loop sum-integrals is 
%
\begin{align*}
     \frac{T^2}{256\pi^4\epsilon}&\bigg[2\times 12\times(0.0625)-2\times 24\times(-0.0625)\nonumber\\
    &-2\times 24\times(0.0078) -2\times 24\times(0.0234)\bigg]\approx 3.002\,,
\end{align*}
%
that cancels the last $-3$ in Eq.~\eqref{eq:cancellations}. (It has to be so because $\delta m_{G_0}^2$ has no $y_t^4 g_s^2$ term.)

Alternatively, we can require full cancellation of poles in all terms of the matched masses to reconstruct the coefficients in Tab.~\ref{tab:master-coefficients} analytically. Thus, the analytic double-pole coefficients in the preceding example
are $1/16$, $-1/16$, $1/128$ and $3/128$, respectively, giving exactly $3$.

The scale dependence of the Higgs and colour Debye masses are shown in Figs.~\ref{fig:scaledependence1} and \ref{fig:scaledependence2}. Numerical uncertainties from sum-integral evaluation have negligible impact on the plots. Interestingly, the finite pieces of the new sum-integrals do not affect the colour Debye mass, as they enter into the matching proportional to $(d-3)$. Something similar was shown also in pure QCD~\cite{Ghisoiu:2015uza}.

\section{Discussion}
We have achieved full $\mathcal{O}(g^6)$ accuracy in the
determination of the hard matching of thermal masses in the SM,
substantially extending previous results. At $T=1\,\mathrm{TeV}$,  the renormalization-scale band of $m_{G_0}^2/T^2$ narrows by approximately $70\%$ relative to the $\mathcal{O}(g^4)$ result. The sum-integral evaluation methods presented here can also be applied to new-physics scenarios. A determination of thermal masses in the SMEFT will be presented in forthcoming work~\footnote{Work in progress with Javier Fuentes-Mart\'in, Luis Gil, Javier L\'opez-Miras and Adri\'an Moreno-S\'anchez.}, together with two-loop quartic matching and automation within \texttt{Matchotter}~\cite{Fuentes-Martin:2026bhr}.




\section*{Acknowledgments} 
I thank Luis Gil for useful discussions. This work is supported by the European Research Council under grant agreement n. 101230200.


\bibliographystyle{apsrev4-2}
\bibliography{references}

\end{document}